\documentclass[runningheads]{llncs}
\usepackage[T1]{fontenc}
\usepackage[dvipdfmx]{graphicx}
\usepackage{amssymb, amsmath, bm}
\usepackage{algorithm,algorithmic}
\usepackage{booktabs}
\usepackage{enumitem}
\usepackage{xcolor}
\usepackage{xurl}
\usepackage[misc,geometry]{ifsym}

\newcommand{\Argmax}{\mathop{\rm argmax}\limits}
\newcommand{\argmax}{\mathop{\rm argmax}\nolimits}

\begin{document}
\title{Efficient Preference Elicitation in Iterative Combinatorial Auctions with Many Participants}
\titlerunning{Efficient Preference Elicitation in ICA with Many Participants}
%
\author{
Ryota Maruo$^{\text{(\Letter)}}$ \and
Hisashi Kashima 
}
%
\authorrunning{Ryota Maruo \and Hisashi Kashima}
%
\institute{Kyoto University, Sakyo-ku, Kyoto, Japan\\
\email{mryota@ml.ist.i.kyoto-u.ac.jp},
\email{kashima@i.kyoto-u.ac.jp}
}

\maketitle              
\begin{abstract}
We study the problem of achieving high efficiency in iterative combinatorial auctions (ICAs).
ICAs are a kind of combinatorial auction where the auctioneer interacts with bidders to gather their valuation information using a limited number of queries, aiming for efficient allocation.
Preference elicitation, a process that incrementally asks bidders to value bundles while refining the outcome allocation, is a commonly used technique in ICAs.
Recently, the integration of machine learning (ML) into ICAs has significantly improved preference elicitation.
This approach employs ML models that match the number of bidders, estimating each bidder's valuation functions based on their reported valuations.
However, most current studies train a separate model for each bidder, which can be inefficient when there are numerous bidders with similar valuation functions and a limited number of available queries.
In this study, we introduce a multi-task learning method to learn valuation functions more efficiently.
Specifically, we propose to share model parameters during training to grasp the intrinsic relationships between valuations.
We assess the performance of our method using a spectrum auction simulator.
The findings demonstrate that our method achieves higher efficiency than existing methods, especially in scenarios with many bidders and items but a limited number of queries.

\end{abstract}
\section{Introduction}
Combinatorial auction (CA) is an effective mechanism for allocation, allowing bidders to place bids on sets of items, known as {\em bundles}.
This approach enables them to express their complex preferences for items, considering the complementarity and substitutability among the items.
These auctions have significant applications in various sectors, such as the sale of spectrum licenses \cite{cramton2013spectrum}, real estate spaces \cite{goossens2014solids}, and airport access rights \cite{BALL2018186}.

The primary challenge in CAs is the exponential number of possible bundles. 
The well-known Vickrey-Clarke-Groves (VCG) mechanism \cite{clarke1971,groves1973,vickrey1961} assumes access to bidders' full valuation functions, that is, bidders need to value all the bundles, which is impractical with a large number of items.
To circumvent this issue, extensive research has focused on preference elicitation \cite{blum2004preference}, an iterative process in which bidders are queried to provide information about their valuations instead of reporting on all possible bundles. 
This approach simplifies the bidding process by reducing the information required from each participant.
The queries for preference elicitation are classified into various types, including value queries, which ask for a valuation of a specific bundle, and demand queries, which inquire about the bundle with maximum utility given prices~\cite{sandholm05chap}.
A notable implementation is the combinatorial clock auction, primarily used in spectrum auctions~\cite{cca10.1111/ecoj.12404}, which employs demand queries to elicit preference information.

Recently, machine learning has enhanced preference elicitation.
Following the seminal works \cite{blum2004preference,Lahaie04appl}, a series of studies have developed under the concept of ML-powered iterative combinatorial auctions (ICAs).
In ML-powered ICAs, machine learning models are employed for each bidder to learn their valuation functions based on queries, with the trained model parameters determining the next queries \cite{brero2017probably}.
Brero et al.~\cite{brero2017probably} first proposed the framework for ML-based ICAs, and subsequent works \cite{weissteiner22monotone,Weissteiner_Heiss_Siems_Seuken_2023,Weissteiner_Seuken_2020} have developed more efficient elicitation algorithms. 
From a theoretical perspective, the worst-case communication complexity is exponential in the number of items when bidders have arbitrary monotone valuation functions \cite{NISAN2006192}.
Nonetheless, preference elicitation can still be effective with machine learning support.

However, most existing studies train models separately for each bidder.
Influenced by the mainstream framework \cite{brero18combinatorial,brero2021machine}, the majority of subsequent works use each ML model to approximate the corresponding bidder's valuation function independently. 
This method, while effective, can lead to inefficiencies, particularly when several bidders have similar valuation functions.

The implications of these inefficiencies are further illustrated in the following examples, which demonstrate the potential overlap in valuation functions across different auction settings.

Possibly many bidders can have similar valuation functions in certain CA applications. 
In auction theory, correlated valuation distributions are a natural assumption often employed to analyze auctions like mineral rights or spectrum auctions~\cite{mcmillan1994spectrum,milgrom1982auction}.
High participation can be observed in real-world spectrum auctions in the US, where the number of bidders and the number of items (licenses) can be more than 50 and 1,000, respectively~\cite{fcc_auction97}.
In addition, auctions in different applications such as logistics can have more than 100 bidders on average \cite{vries03comb}.
These two phenomena combined—similar valuation strategies among bidders and a high participation rate—suggest that, in scenarios where both conditions apply, the implications for auction design are significant. 

When there are many bidders with similar valuation functions, avoiding repetitive queries to different bidders is crucial to reduce time costs; however, separately trained models cannot detect the similarity and hence will necessitate the repetition of the same queries across bidders. 

In scenarios with several bidders whose valuations are somewhat related, there is a clear need for a methodology for handling these complexities, aiming for efficient allocation with fewer queries. 
In this context, joint model training emerges as a crucial method.
Jointly trained models are expected to capture the intrinsic relatedness of valuation functions more effectively by sharing information across models.
Despite its potential, this type of approach is underexplored in existing studies like Weissteiner et al.~\cite{weissteiner22monotone,Weissteiner_Heiss_Siems_Seuken_2023}, which instead invent several better model architectures to make more efficient preference elicitation algorithms.

In this paper, we propose a {\em multi-tasking machine learning-powered combinatorial auction} ({\em MT-MLCA}) that integrates multi-task learning into existing machine learning-based preference elicitation algorithms. 
This approach aims to utilize shared information across different bidders to improve efficiency.
Technically, we apply the soft-parameter sharing across models to capture the valuation similarities.
In addition, we incorporate a method using bidders' ID features to assist in differentiating between tasks.

We experimentally assess the effectiveness of multi-task learning in improving the existing methods.
We conduct evaluations in experimental settings characterized by a large number of bidders but limited availability of queries. 
The experimental results show that our multi-tasking approach yields higher efficiency than the existing method under 196 items and over 30 bidders, or 50 bidders with similar valuation functions and 98 or 196 items.

Our contributions are summarized as follows:
\begin{itemize}
    \item We introduce soft-parameter sharing from the realm of multi-task learning to ML-powered ICAs, improving efficiency by using shared bidder information.

    \item We validate our approach through experiments in scenarios with many bidders and limited queries, demonstrating its practical advantages.
\end{itemize}
Our code is available at 
\url{https://github.com/MandR1215/EfficientPE}

\section{Related Work}
\subsection{Preference Elicitation in Combinatorial Auctions}
Combinatorial auctions present several practical challenges, such as the difficulty for bidders to report complete valuations for the exponentially large number of possible bundles. 
This challenge has motivated researchers to explore preference elicitation, defined as ``the process of asking questions about the preferences of bidders to best divide some set of goods'' \cite{blum2004preference}. 
The initial framework for this approach was proposed by Conen and Sandholm~\cite{conen01preference}, and the general procedure can be described as follows~\cite{sandholm05chap}: 
\begin{enumerate}
    \item Initialize $t := 0$ and let $C_0$ represent the prior information available to the auctioneer about the bidders' valuation functions.
    \item Given the current information on valuation functions $C_t$, decide whether to
        \begin{enumerate}
            \item terminate the process and determine an allocation and payments, or
            \item choose a set of queries $Q_t$, gather responses to these queries from the bidders, and update the current information from $C_{t}$ to $C_{t+1}$, incrementing $t$ to $t+1$.
        \end{enumerate}
\end{enumerate}

More recent studies \cite{brero2017probably,brero18combinatorial,weissteiner22monotone,Weissteiner_Heiss_Siems_Seuken_2023,Weissteiner_Seuken_2020}, including this study, can be viewed as instances of the above general framework.

\subsection{Automated Mechanism Designs in Auctions}
Conitzer and Sandholm~\cite{conitzer2002complexity} first introduced the concept of {\em automated mechanism design (AMD)}, studying the computational complexity involved in designing direct revelation mechanisms.
This foundational work spurred a series of studies focused on the automatic design of revenue-optimal combinatorial auctions \cite{balcan2016sample,conitzer2004self,curry2023differentiable,dutting2019optimal,likhodedov2004methods}.

Machine-learning techniques have significantly advanced the domain of auction mechanisms. Initially, AMD approaches relied on linear programming, which proved computationally infeasible for larger scales \cite{guo2010computationally}.
To make AMD further scalable, D\"{u}tting et al.~\cite{dutting2019optimal} introduced RegretNet, a deep learning model designed to learn revenue-maximizing auction mechanisms. 
RegretNet aims to maximize revenue while minimizing {\em regret}, which measures the extent of deviation from strategy-proofness.
Conversely, D\"{u}tting et al.~\cite{dutting21optimal} developed RochetNet, a model specifically constructed to adhere strictly to strategy-proofness, as opposed to RegretNet, which only approximately achieves this. Initially, RochetNet was limited to scenarios with a single bidder. 
However, recent work by Curry et al.~\cite{curry2023differentiable} has expanded its application to multiple bidders, employing an affine maximizer auction framework.

While most existing studies in this area often assume that bidders have additive utilities, our research considers general combinatorial utilities, aligning with the prevailing trend, as exemplified by related studies \cite{weissteiner22monotone,Weissteiner_Heiss_Siems_Seuken_2023,Weissteiner_Seuken_2020}.

\section{Preliminaries}
We first describe the problem setting and then introduce the ML-powered ICA proposed by Brero et al.~\cite{brero2021machine}. 

\subsection{Problem Setting}
For any $k\in\mathbb{N}$, let $[k]$ denote the set of $\{1,\dots, k\}$ and we denote by $\mathbb{R}_{\ge 0}$ the set of non-negative real numbers.

Consider a CA setting where we are given a set of bidders $N := [n]$ and a set of items $M:= [m]$.
A {\em bundle} is a subset of $M$, denoted by a vector $\bm{x}\in\{0,1\}^m$, where $x_k = 1$ if and only if the $k$-th item belongs to the bundle.
Each bidder $i$ has a private valuation function $v_i:\{0,1\}^m\to\mathbb{R}_{\ge 0}$ which gives $i$'s true value for a bundle.
We assume that several bidders have similar valuation functions.
This assumption corresponds to a correlated valuation distributions over bidders as discussed in Milgrom and Weber~\cite{milgrom1982auction}, which does not imply that the auctioneer can observe before starting the auction that their true valuations are correlated.

The ICA mechanism outputs an allocation of bundles and monetary payments.
We denote by $A := [\bm{a}_1~\cdots~\bm{a}_n] \in \{0,1\}^{m\times n}$ an allocation of items to bidders, where the $i$-th column $\bm{a}_i := (a_{1i},\cdots, a_{mi})^\top\in\{0,1\}^m$ is the bundle which the bidder $i$ obtains.
An allocation $A$ is {\em feasible} if $\sum_{i\in N} a_{ji} \le 1$ for all $j \in M$.
We let $\mathcal{F} := \{A\in \{0,1\}^{m\times n}~|~\sum_{i\in N} a_{ji} \le 1,~\forall j \in M\}$ be the set of all feasible allocations.
Payments are denoted by a vector $\bm{p} := (p_1, \dots, p_n)^\top\in\mathbb{R}^n_{\ge 0}$. 
We assume bidders have quasi-linear utilities $u_i(\bm{a}_i, \bm{p}) := v_i(\bm{a}_i) - p_i$ for an allocation $A$ and payments $\bm{p} = (p_1,\dots, p_n)^\top\in\mathbb{R}^n_{\ge 0}$.
Simultaneously, the auctioneer receives utility $u_\mathrm{auctioneer}(\bm{p}) := \sum_{i\in N} p_i$.
Given a feasible allocation $A\in\mathcal{F}$, the {\em social welfare} is defined as $V(A) := \sum_{i\in N} v_i(a_i)$.
This is equal to the sum of utilities of the auctioneer and bidders for any payments $\bm{p}$ because $\sum_{i\in N} u_i(\bm{a}_i,\bm{p}) + u_\mathrm{auctioneer}(\bm{p}) = \sum_{i\in N} (v_i(\bm{a}_i) - p_i) + \sum_{i\in N} p_i = \sum_{i\in N} v_i(\bm{a}_i) = V(A)$.
An {\em efficient} allocation is the social-welfare maximizing allocation $A^* := \argmax_{A\in\mathcal{F}} V(A)$.
For a given feasible allocation $A\in\mathcal{F}$, the {\em efficiency} is defined as $V(A) / V(A^*)$. 
Higher efficiency indicates larger social welfare.

An ICA mechanism aims to find an approximately efficient allocation.
During the procedure, the mechanism repeatedly asks bidders to report their valuation on some bundles to determine a final allocation.
Let $\bar{v}_i(\bm{x})$ be the possibly untruthful reported valuation for a bundle $\bm{x}$.
We assume that $\bar{v}_i(\bm{x})\ge 0$ for all $\bm{x}\in\{0,1\}^m$.
The reported bundle-value pairs are denoted by the set of $R_i := \{(\bm{x}^{(k)}, \bar{v}_i(\bm{x}^{(k)})\}_{k = 1}^{n_i}$, where $n_i$ denotes the number of totally reported pairs.
Given $R := (R_1,\dots, R_n)$, the {\em reported social welfare} for an allocation $A\in\mathcal{F}$ is defined as $\overline{V}(A|R) := \sum_{i\in N: (\bm{a}_i, \bar{v}_i(\bm{a}_i))\in R_i} \bar{v}_i(\bm{a}_i)$, that is, the sum of reported valuations on bundles contained both in $A$ and $R$.
The final allocation is the allocation that maximizes the reported social welfare, which is determined by
\begin{align}
    A^*_R := \Argmax_{A\in \mathcal{F}} \overline{V}(A|R). \label{eq:repalloc}
\end{align}
The objective is to collect bundle-value pairs for the final allocation $A^*_R$ to be as efficient as possible \cite{Weissteiner_Seuken_2020}. 
Formally, given a constraint on the maximum number of queries per bidder, denoted as $c_e\in\mathbb{N}$, we aim to determine a set $R$ such that:
$
    R\in \argmax_{R: |R_i| \le c_e} {V(A^*_R)}/{V(A^*)}. 
$

In practice, the maximum query cap $c_e$ must be small to reduce consideration costs on bidders, while keeping the efficiency $V(A^*_R)/V(A^*)$ as high as possible.
According to Weissteiner et al.~\cite{weissteiner22monotone}, even a minor increase in this efficiency—just 1 percentage point—can result in substantial monetary gains, potentially amounting to hundreds of millions of dollars.
Therefore, achieving higher efficiency as well as reducing the number of queries not only streamlines the auction process but also enhances potential revenue, so the goal here is to optimize with the aim of maximizing efficiency.

\subsection{ML-powered ICA}
\begin{algorithm}[tb]
    \caption{\textsc{MLCA}~\protect\cite{brero2021machine}}
    \label{alg:mlca}
    \textbf{Input}: Numbers of queries $Q^\mathrm{init}, Q^\mathrm{max}, Q^\mathrm{round}$\\
    \textbf{Output}: \makebox[0.4\linewidth][l]{Allocation $A^*_R$ and payments $\bm{p}(R)$}
    \begin{algorithmic}[1] 
        \FOR {$i\in N$}
            \STATE Receive reports $R_i$ for $Q^\mathrm{init}$ randomly drawn bundles
        \ENDFOR
        \FOR{$k = 1,\dots, \lfloor (Q^\mathrm{max}-Q^\mathrm{init})/Q^\mathrm{round}\rfloor$}
            \STATE Initialize new queries $q^\mathrm{new}_i = \emptyset$ for each $i\in N$
            \FOR{$i\in N$}
                \STATE Draw uniformly without replacement $(Q^\mathrm{round}-1)$ bidders from $N\setminus \{i\}$ and store them in $\tilde{N}$\label{algline:marginal_econ_6}
                \FOR{$j\in \tilde{N}$}
                    \STATE $q^\mathrm{new}_i = q^\mathrm{new}_i\cup \textsc{NextQueries}(N\setminus\{j\}, R_{-j})_i$\hfill $\triangleright$ Add to only $i$'s new query \label{algline:marginal_econ_8}
                \ENDFOR
            \ENDFOR
            \FOR{$i\in N$}
                \STATE $q^\mathrm{new}_i = q^\mathrm{new}_i\cup\textsc{NextQueries}(N,R)_i$ \label{algline:main_econ}
            \ENDFOR
            \FOR{$i\in N$}
                \STATE Receive reports $R_i^\mathrm{new}$ for $q_i^\mathrm{new}$, set $R_i = R_i\cup R_i^\mathrm{new}$
            \ENDFOR
        \ENDFOR
        \STATE Compute $A^*_R$ as in \eqref{eq:repalloc}
        \STATE Compute VCG payments $\bm{p}(R)$ as in \eqref{eq:vcgpayment}
        \RETURN Allocation $A^*_R$ and payments $\bm{p}(R)$
    \end{algorithmic}
\end{algorithm}

\begin{algorithm}[tb]
    \caption{\textsc{NextQueries}~\protect\cite{brero2021machine}}
    \label{alg:nxtqry}
    \textbf{Input}: Subset of bidders $I$ and reported bundle-value pairs $R$\\
    \textbf{Params}: ML algorithm $\mathcal{A} = (\mathcal{A}_i)_{i\in N}$\\
    \textbf{Output}: \makebox[0.4\linewidth][l]{New query profile $\mathcal{Q} = [\bm{q}_i]_{i\in N}$}
    \begin{algorithmic}[1] 
        \FOR {$i\in I$}
            \STATE Fit $\mathcal{A}_i$ on $R_i$ and obtain $\mathcal{A}_i[R_i]$\hfill $\triangleright$ Estimation step\label{algline:estimation_step}
        \ENDFOR
        \STATE Solve $\mathcal{Q} = [\bm{q}_1~\cdots~\bm{q}_n]\in \argmax_{A\in\mathcal{F}}\sum_{i\in I}\mathcal{A}_i[R_i](\bm{a}_i)$
        \item[] \hfill $\triangleright$ Optimization step\label{algline:optimization_step}
        \FOR{$i\in I$}
            \IF{$(\bm{q}_i, \bar{v}_i(\bm{q}_i)) \in R_i$}
                \STATE \makebox[0.4\linewidth][l]{Define $\mathcal{F}' := \{A\in\mathcal{F}~|~\bm{a}_i\neq \bm{x}, \forall(\bm{x}, \bar{v}_i(\bm{x}))\in R_i\}$}
                \STATE Resolve $\mathcal{Q}' \in \argmax_{A\in\mathcal{F}'}\sum_{l\in I}\mathcal{A}_l[R_l](\bm{a}_l)$
                \STATE Update $\bm{q}_i$ to $\bm{q}_i'$, the $i$-th column of $\mathcal{Q}'$
            \ENDIF
        \ENDFOR
        \RETURN $\mathcal{Q} = [\bm{q}_1~\dots~\bm{q}_n]$
    \end{algorithmic}
\end{algorithm}

We describe the machine learning-powered combinatorial auction (\textsc{MLCA}) in Algorithm \ref{alg:mlca} for the machine learning-powered ICAs proposed by Brero et al.~\cite{brero2021machine} with slightly changed notations from Weissteiner et al.~\cite{weissteiner22monotone}.

\textsc{MLCA} proceeds in rounds by repeatedly asking valuations for specific bundles until reaching the maximum round $Q^\mathrm{max} = c_e$. 
During the procedure, \textsc{NextQueries} in Algorithm \ref{alg:nxtqry} is invoked to compute the next queried bundles. 
This computation involves two key steps: at the estimation step (Line \ref{algline:estimation_step}), a machine learning model $\mathcal{A}_i: \{0,1\}^m\to\mathbb{R}_{\ge 0}$ is trained on each bidder $i$'s reported bundle-value pairs $R_i$ through regression to estimate the bidders' valuation function. 
Subsequently, at the optimization step (Line \ref{algline:optimization_step}), the most promising allocation $\mathcal{Q}$ is calculated based on the estimated valuation functions. 
Finally, the next queries are calculated after the exclusion of previously asked bundles and re-computation of tentative allocations. 
\textsc{MLCA} guarantees that truthful bidding is an ex-post Nash equilibrium under several assumptions \cite{brero2021machine}. 
Our method preserves this property, which we discuss in \ref{apdx:incentive}.

\textsc{MLCA} outputs the final allocation, denoted by $A^*_R$, and the payment vector $\bm{p}(R)$.
The allocation $A^*_R$ is obtained by solving equation \eqref{eq:repalloc}. 
The payment vector $\bm{p}(R) = (p(R)_i)_{i\in N}$ represents the VCG Payments, calculated in a manner akin to the original VCG rule. 
Specifically, let $R_{-i} := (R_1,\dots, R_{i-1}, R_{i+1}, \dots, R_{n})$ denote the tuple of reported bundle-value pairs except for bidder $i$'s one, and \textsc{MLCA} calculates the payments by
\begin{align}
    p(R)_i := \sum_{j\in N\setminus\{i\}} \bar{v}_j(\bm{a}^*_{R_{-i},j}) - \sum_{j\in N\setminus\{i\}} \bar{v}_j(\bm{a}^*_{R,j}),
    \label{eq:vcgpayment}
\end{align}
where $\bm{a}^*_{R,j}$ denotes the $j$-th column of $A^*_R$, and $A^*_{R_{-i}} := [\bm{a}^*_{R_{-i}, 1}~\cdots~\bm{a}^*_{R_{-i}, n}]$ is defined as 
$
    A^*_{R_{-i}} := \argmax_{A\in\mathcal{F}} \sum_{j\in N\setminus \{i\}: (\bm{a}_j, \bar{v}_j(\bm{a}_j))\in R_j} \bar{v}_j(\bm{a}_j),
$
which represents the allocation that maximizes the reported social welfare, excluding the contribution of bidder $i$.

In the estimation step, each model $\mathcal{A}_i: \{0,1\}^m\to\mathbb{R}_{\ge 0}$ estimates bidder $i$'s valuation function via labeled data $R_i = \{(\bm{x}_i^{(k)}, \bar{v}_i(\bm{x}_i^{(k)}))\}_k$.
Several studies have investigated ML models $\mathcal{A}$ in Algorithm \ref{alg:nxtqry}, including SVMs \cite{brero2017probably} and deep learning models \cite{weissteiner22monotone,Weissteiner_Heiss_Siems_Seuken_2023,Weissteiner_Seuken_2020}. 
Here, we assume the monotone-value neural network (MVNN) proposed in Weissteiner et al.~\cite{weissteiner22monotone}; however, our proposed method can be extended to other architectures.
The MVNN, a multi-layer perceptron, is constructed as a monotone set function. Formally, an MVNN $\mathcal{A}_i = \mathcal{N}_i(\bm{W}^i, \bm{b}^i)$ has $(K_i-1)$ hidden layers with non-negative weight matrices $\bm{W}^i = (W^{i, 1}, \dots, W^{i, K_i})\ge 0$, non-positive biases $\bm{b}^i = (\bm{b}^{i,1},\dots, \bm{b}^{i, K_i-1})\le 0$, and the bounded ReLU activation function $\varphi_{0,t}(z) := \min(t, \max(0, z))$.
An MVNN satisfies monotonicity, i.e., for all $\bm{x}, \bm{y} \in \{0,1\}^m$, if regarding $\bm{x}$ and $\bm{y}$ as subsets $\bm{x}, \bm{y}\subseteq \{0,1\}^m$ and set inclusion holds $\bm{x}\subseteq \bm{y}$, then $\mathcal{N}_i(\bm{W}^i, \bm{b}^i)(\bm{x})\le \mathcal{N}_i(\bm{W}^i, \bm{b}^i)(\bm{y})$.
The model $\mathcal{A}_i$ trained on $R_i$ is denoted by $\mathcal{A}_i[R_i]$.

The optimization step generates the allocation that gives the largest social welfare with respect to the estimated valuation ${\mathcal{A}_i[R_i]}_{i\in I}$. 
Because $I = N$ or $I = N\setminus\{j\}$ as described in Line \ref{algline:marginal_econ_8} and \ref{algline:main_econ} of Algorithm \ref{alg:mlca}, the bundles allocated to bidders in $N\setminus I$ is uniquely determined.
This step is implemented and solved via a mixed integer linear programming which is written by the model parameters~\cite{weissteiner22monotone}.

\section{Proposed Method}
We propose our MT-MLCA, the integration of a multi-task learning approach with MLCA and MVNN.
Initially, we will outline the process of sharing valuation information among bidders. 
Subsequently, we will present our technique, which utilizes bidders' ID features to distinguish and capture task differences.

\subsection{Soft Parameter-Sharing}
In the estimation step in Algorithm \ref{alg:nxtqry}, the models $\mathcal{A}_i$ are trained exclusively on their respective datasets $R_i$, but this approach is ineffective when multiple bidders have similar valuation functions.
Instead, we train $\{\mathcal{A}_i\}_{i\in I}$ on $\{R_i\}_{i\in I}$ simultaneously by applying multi-task learning. 
Technically, we adopt a simple soft parameter-sharing approach proposed by Duong et al.~\cite{duong-etal-2015-low}.
Our multi-tasking approach preserves the incentive properties of \textsc{MLCA} as discussed in \ref{apdx:incentive}.

We focus on the estimation step when \textsc{NextQueries} is invoked for a set of bidders $I$.
Let $R_I := (R_i)_{i\in I}$ represent the sets of reported bundle-value pairs, and $(\mathcal{N}_i(\bm{W}^i, \bm{b}^i))_{i\in I}$ denote the MVNNs participating in the estimation step.
We assume that all MVNNs have an identical architecture with $(K-1)$ hidden layers, which we discuss in \ref{apdx:identical}.

In our soft parameter-sharing approach, we aim to minimize regression loss while maintaining proximity among the models.
Let $\hat{y}_i^{(k)} := \mathcal{N}_i(\bm{W}^i, \bm{b}^i)(\bm{x}_i^{(k)})$ denote the estimation of $y_i^{(k)} := \bar{v}_i(\bm{x}_i^{(k)})$ given $(\bm{x}_i^{(k)}, \bar{v}_i(\bm{x}_i^{(k)}))\in R_i$.
We train $(\mathcal{N}_i(\bm{W}^i, \bm{b}^i))_{i\in I}$ by minimizing:
\begin{align}
    &\mathrm{loss}(\{(\bm{W}^i, \bm{b}^i)\}_{i\in I}) := \sum_{i\in I} \sum_{k = 1}^{|R_i|} L(y^{(k)}_i, \hat{y}^{(k)}_i) + \lambda \sum_{i, j\in I} \sum_{s\in S}\|W^{i, s} - W^{j, s}\|_\mathrm{F}^2,
    \label{eq:softloss}
\end{align}
where $L$ is the regression loss (e.g., $\ell$2-loss), $\lambda$ is a hyperparameter, $S\subset [K]$ is the indices of shared weights, and $\|X\|_\mathrm{F} = \sqrt{\mathrm{tr}({X^\top X})}$ is the Frobenius norm of a matrix $X$. 
The parameters indexed by any $s\in [K]\setminus S$ contribute to task-specific components.

For our experimental evaluation, we explore two configurations of the sharing indices $S$.
The first configuration, referred to as {\em MT-MLCA-F}, includes indices in the range $S = {1,\dots, \lfloor K/2 \rfloor}$. 
The second configuration, named {\em MT-MLCA-R}, encompasses indices from ${\lfloor K/2 \rfloor,\dots, K}$. 
Here, `F' in MT-MLCA-F represents the `Front' part of the range, and `R' in MT-MLCA-R.
Note that both the MT-MLCA-F and MT-MLCA-R configurations incorporate the ID injection technique described in the following section.

\subsection{ID Injection}
We utilize bidder IDs to assist models in distinguishing task differences.
Conceptually, we make a feature vector for each bidder ID $i\in N$ and incorporate it into the corresponding model $\mathcal{A}_i$.
In our experiments, using this technique enhanced the performance of MT-MLCA-F and MT-MLCA-R, enhancing the benefits of multi-task learning.

A bidder ID $i\in N$ is crucial for the model's ability to capture task differences and adapt to individual bidder characteristics.
The MVNN, by construction, is assumed to have an input dimension equal to the number of items.
Consequently, we cannot input the concatenation of a bundle vector with an ID representation directly.
Hence, we propose to inject an ID feature into a hidden layer output to adapt an MVNN for individual bidder ID recognition.

Our ID injection is as follows:
We make a trainable embedding vector $\bm{e}^i\in \mathbb{R}^d_{\le 0}$ that represents the $d$-dimensional non-positive embedding for an ID $i\in N$.
MVNN assumes that all the bias terms must be non-positive elements, so our embedding must exist in $\mathbb{R}^d_{\le 0}$.
We then inject $\bm{e}^i$ into the $j$-th layer, altering the layer's parameters $W^{i,j}$ and $\bm{b}^{i,j}$ as:
\begin{align}
    W^{i,j} := \begin{bmatrix}
        W^{i,j}\\
        O
    \end{bmatrix}, \quad
    \bm{b}^{i,j} := \begin{bmatrix}
        \bm{b}^{i,j}\\
        \bm{e}^i
    \end{bmatrix},
\end{align}
where $O$ is the zero-matrix with appropriate dimensions.
This modification does not change the input dimension because the number of columns of $W^{i,j}$ and $\bm{b}^{i,j}$ remains the same.

\section{Experimental Result}
In this section, we conduct simulation-based experiments.
First, we briefly summarize the evaluation simulator.
We then describe the experimental setup and results, followed by a discussion.

\subsection{Spectrum Auction Test Suite (SATS)}
We employ the Spectrum Auction Test Suite (SATS) \cite{weiss17sats} to generate combinatorial auction instances.
SATS generates combinatorial auction instances using a {\em value model}—an analytic or algorithmic representation of a bidder's valuation function, which is realized by varying random parameters \cite{weiss17sats}.

Our study primarily focuses on the Multi-Region Value Model (MRVM).
In MRVM, bidders are mobile network operators, and items are the licenses for using spectrum band blocks over one region out of 14 regions.
Bidders are classified into three kinds: local, regional, and national. Note that each kind of bidder shares the structure of value functions except for random parameters.
Those varieties are described as follows:
\begin{itemize}
    \item 
        A local bidder is interested in specific regions and has a positive value on a license for them. 
        The interested regions are drawn uniformly from the set of all the regions.
    \item 
        A regional bidder has a headquarters in one region. 
        This kind of bidder values a bundle considering the distance of its licenses from the headquarters.
        The location of the headquarters is drawn uniformly.
    \item 
        A national bidder prefers to cover as many regions as possible and, therefore, has higher values on bundles that contain licenses distributed over nearly all the regions.
\end{itemize}

MRVM contains 3 local bidders, 4 regional bidders, and 3 national bidders, along with 98 items by default, but can be modified to generate differently-sized auction problems.
When using SATS, we follow the prior work \cite{weissteiner22monotone} and assume truthful bidding i.e., $\bar{v}_i = v_i$. 
This assumption is justified within our auction design, as it maintains the incentive properties outlined in \ref{apdx:incentive}.

\subsection{Experimental Setup}
We evaluate the performance of our multi-tasking on large-scale and small-data settings.

\subsubsection{MLCA Configuration}
For \textsc{MLCA}, we set $Q^\mathrm{init} = Q^\mathrm{round} = 1$ and $Q^\mathrm{max} = 10$ to realize small-data settings. 

In the MRVM auction instances, we modify the number of bidders in two primary settings. These are (1) multiplying the default number by 1 to 5 times, which results in $(3l, 4l, 3l), l \in \{1,2,3,4,5\}$, and (2) having 50 bidders of the same kind, represented as $(50, 0, 0)$, $(0, 50, 0)$, or $(0, 0, 50)$.

These adjustments are applied to local, regional, and national bidders, respectively.
The former settings serve to show the adaptability of our approach to varying problem sizes. 
Conversely, the latter settings are designed to demonstrate its effectiveness in scenarios where task characteristics are common. 
We conduct 10 auction instances for each of these settings.

Unlike existing studies \cite{brero2021machine,weissteiner22monotone,Weissteiner_Heiss_Siems_Seuken_2023,Weissteiner_Seuken_2020}, we also consider two settings for the number of items: 98 and 196.
We yielded the setting for $m=196$ by doubling the spectrum band block quantities to increase the number of items effectively.

The above settings change the default settings of MRVM in SATS, but we argue that this is not unrealistic.
MRVM models spectrum auctions ``in which multiple frequency bands as well as geographic
complementarities determine the bidders’ values in large
real-world auctions such as in the US and Canada.''~\cite{weiss17sats}
This means that MRVM is not limited to the default setting, but can be used generally as a benchmark that models such auction environments. 
Furthermore, in real-world spectrum auctions in the US, the number of bidders and the number of items (licenses) can exceed 50 and 1,000, respectively~\cite{fcc_auction97}.
In summary, there are real-world applications where the MRVM model can be applied as well as the number of bidders are more than the default settings.
Based on these facts, we argue that our experiments have practical implications.

\subsubsection{Model Architectures}
Given our assumption that bidder types are indiscernible, we use the same MVNN architecture for all bidders rather than using hyper-parameter optimized configurations tailored to each bidder kind. 
We adopt the architecture designed for local bidders in the previous research \cite{weissteiner22monotone}.

For the shared parameter sets $S$ in our multi-task learning, we consider two settings: $S = \{1,\dots, \lfloor K/2\rfloor\}$ for MT-MLCA-F, and $S = \{\lfloor K/2\rfloor,\dots, K\}$ for MT-MLCA-R.
$K$ is given in the model configuration for local bidders (see Weissteiner et al.~\cite{weissteiner22monotone} for more details).
MT-MLCA-F first adopts feature extraction in a shared way and then independently solves regressions for different bidders.
Conversely, MT-MLCA-R extracts features independently and then utilizes them for a shared regression.
We set the regularization factor $\lambda = 10^{-10}$, and the regression loss function $L$ be the mean squared error loss as in the setting for local bidders \cite{weissteiner22monotone}.
For the ID injection, 
we use $d=4$ dimensional embeddings and inject it to the layer at depth $j=1$.

\subsection{Evaluation Metrics}
We evaluate our methods from two metrics: the efficiency and the mean absolute percentage error (MAPE).

We define efficiency as $V(A^*_R)/V(A^*)$. 
Here, $A^*_R$ denotes the final allocation derived from the reported bundle-value pairs $R$, and $A^*$ represents the optimal allocation that SATS provides.
This metric ranges from $0$ to $1$, with higher values indicating better efficiency, 
which means greater social welfare.
We compare the results between \textsc{MLCA} without multi-tasking and \textsc{MT-MLCA-F} or \textsc{MT-MLCA-R} using the one-tailed Wilcoxon signed-rank test on efficiency scores.

We measure the Mean Absolute Percentage Error (MAPE) to evaluate the regression accuracy.
In SATS, national bidders are designed to bid higher values compared to local and regional bidders.
MAPE is particularly appropriate for MRVM, as it allows for the normalization of differences in the bidding scale.

MAPE is calculated as:
\begin{align}
    \mathrm{MAPE} := \frac{1}{n} \sum_{i=1}^n \frac{1}{n_\mathrm{test}} \sum_{l=1}^{n_\mathrm{test}} \left| \frac{\hat{y}^{(l)}_i - y^{(l)}_i}{y^{(l)}_i} \right|\label{eq:mape},
\end{align}
where, $n$ is the number of bidders, $n_\mathrm{test}$ is the number of test bundles, $y^{(l)}_i = v_i(\bm{x}^{(l)})$ is the true valuation, and $\hat{y}^{(l)}_i = \mathcal{N}_i(\bm{W}^i, \bm{b}^i)(\bm{x}^{(l)})$ is the estimated value.
For consistency, a standardized set of test data \(\{\bm{x}^{(l)} \mid l=1,\dots,n_\mathrm{test}\}\) is used across all 10 instances within a single setting.

\subsection{Results}
\begin{figure}[tb]
  \centering
  \begin{minipage}[b]{0.4\linewidth}
    \includegraphics[width=\linewidth]{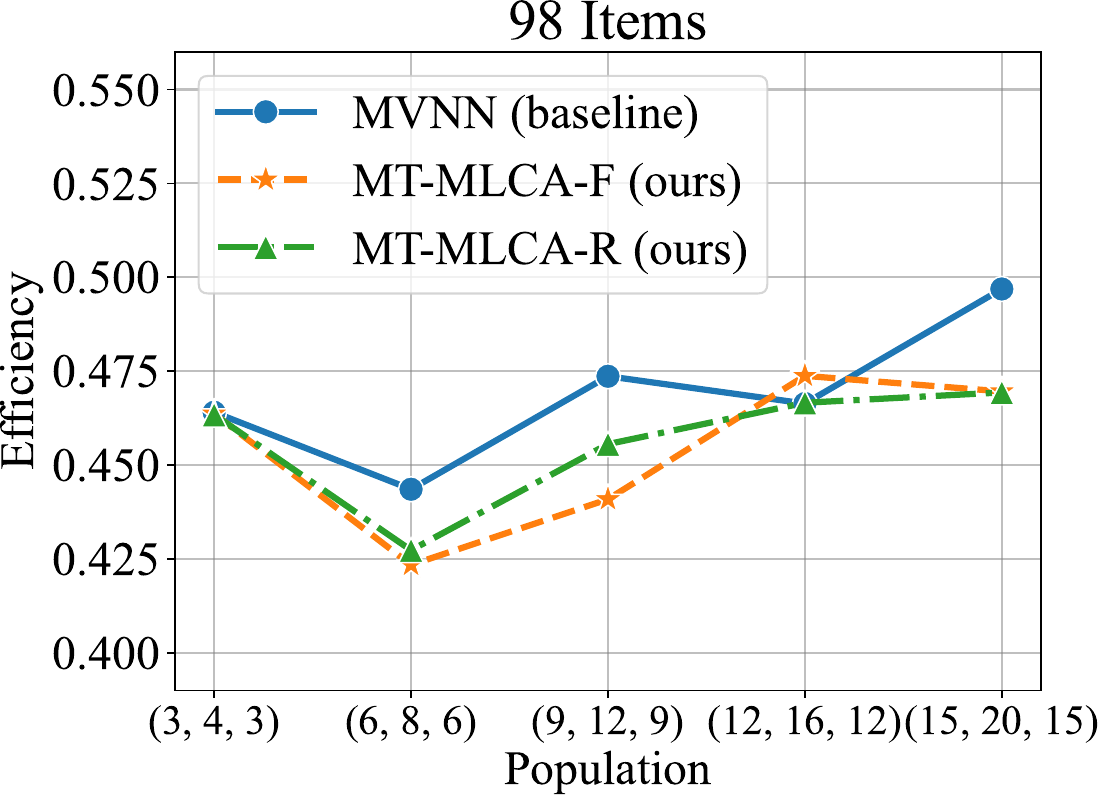}
  \end{minipage}
  \begin{minipage}[b]{0.4\linewidth}
    \includegraphics[width=\linewidth]{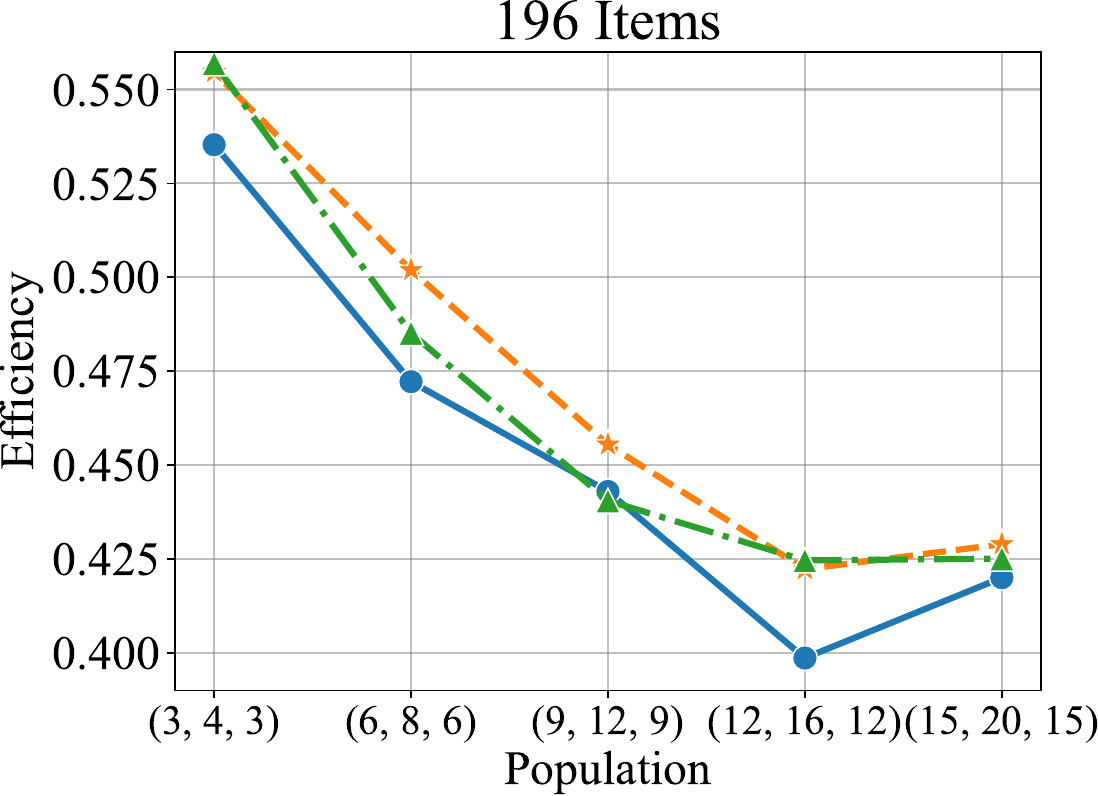}
  \end{minipage}
  \caption{
    Mean efficiency in the $(3l, 4l, 3l) (l=1,\dots, 5)$ settings.
    The horizontal axis represents the number of (local, regional, national) bidders, respectively, while the vertical axis shows the efficiency.
    The left panel displays the efficiency results for 98 items and the right panel for 196 items.
    Both figures share the same legend.
    }
  \label{fig:result_eff_hetero}
\end{figure}

\begin{table*}[tb]
    \centering
    \caption{
        Efficiency results of all the settings. 
        The left four columns represent the number of items and bidders. 
        ``\#L", ``\#R", and ``\#N" denote the number of local, regional, and national bidders, respectively.
        The right three columns show the efficiency of different methods with means and standard deviations. 
        ``MVNN" means \textsc{MLCA} with the plain MVNN without multi-tasking, ``MT-MLCA-F" means MLCA with MVNNs having shared parameters $S=\{1,\dots,\lfloor K/2\rfloor\}$, and ``MT-MLCA-R" means MLCA with MVNNs that share parameters $S=\{\lfloor K/2 \rfloor,\dots, K\}$. 
        Scores with the highest averages are highlighted in bold.
        $\dagger$ signifies the statistical test results.}
    \label{tab:result_eff}
    \scalebox{0.98}{
    \begin{tabular}{rrrrrrr}
        \toprule
        & \multicolumn{3}{c}{\bf \# Bidders} & \multicolumn{3}{c}{{\bf Efficiency} ($\dagger p<0.05$)} \\\cmidrule(lr){2-4}\cmidrule(lr){5-7}
        \multicolumn{1}{c}{\bf \# Items} & \multicolumn{1}{c}{\# L} & \multicolumn{1}{c}{\# R} & \multicolumn{1}{c}{\# N} & \multicolumn{1}{c}{MVNN (baseline)} & \multicolumn{1}{c}{MT-MLCA-F (ours)} & \multicolumn{1}{c}{MT-MLCA-R (ours)} \\\cmidrule(lr){1-1}\cmidrule(lr){2-4}\cmidrule(lr){5-7}
        98  &  3 &  4 &  3 &$\textbf{0.464}\pm\textbf{.0671}$&$\textbf{0.464}\pm\textbf{.0683}$&$0.463\pm.0711$\\
            &  6 &  8 &  6 & $\textbf{0.444}\pm\textbf{.0186}$&$0.423\pm.0266$&$0.427\pm.0397$\\
            &  9 & 12 &  9 & $\textbf{0.474}\pm\textbf{.0535}$&$0.441\pm.0467$&$0.456\pm.0424$\\
            & 12 & 16 & 12 & $0.466\pm.0339$&$\textbf{0.474}\pm\textbf{.0325}$&$0.466\pm.0383$\\
            & 15 & 20 & 15 & $\textbf{0.497}\pm\textbf{.0556}$&$0.470\pm.0337$&$0.469\pm.0497$\\\cmidrule(lr){2-7}
            & 50 &  0 &  0 &$0.603\pm.0289$&$0.620\pm.0576$&$\textbf{0.629}\pm\textbf{.0508}$\\
            &  0 & 50 &  0 &$0.494\pm.0446$&$\textbf{0.516}\pm\textbf{.0598}^\dagger$&$0.515\pm.0702$\\
            &  0 &  0 & 50 &$0.779\pm.0287$&$0.779\pm.0216$&$\textbf{0.781}\pm\textbf{.0273}$\\\midrule
        196 &  3 &  4 &  3 &$0.535\pm.0919$&$0.555\pm.0666$&$\textbf{0.557}\pm\textbf{.0628}$\\
            &  6 &  8 &  6 &$0.472\pm.0448$&$\textbf{0.502}\pm\textbf{.0377}^\dagger$&$0.485\pm.0377$ \\
            &  9 & 12 &  9 &$0.443\pm.0389$&$\textbf{0.455}\pm\textbf{.0430}$&$0.441\pm.0350$\\
            & 12 & 16 & 12 &$0.399\pm.0471$&$0.422\pm.0393$&$\textbf{0.425}\pm\textbf{.0426}^\dagger$\\
            & 15 & 20 & 15 &$0.420\pm.0274$&$\textbf{0.428}\pm\textbf{.0616}$&$0.425\pm.0470$\\\cmidrule(lr){2-7}
            & 50 &  0 &  0 &$0.468\pm.0189$&$\textbf{0.509}\pm\textbf{.0319}^\dagger$&$0.495\pm.0284$\\
            &  0 & 50 &  0 &$\textbf{0.475}\pm\textbf{.0345}$&$0.468\pm.0340$&$\textbf{0.475}\pm\textbf{.0376}$\\
            &  0 &  0 & 50 &$0.728\pm.0164$&$\textbf{0.761}\pm\textbf{.0169}^\dagger$&$0.758\pm.0136$\\\bottomrule
    \end{tabular}
    }
\end{table*}
We present efficiency results in the above settings in Table \ref{tab:result_eff} and visualize the average results of the first settings in Figure \ref{fig:result_eff_hetero}.
Table \ref{tab:result_eff} shows that, on average, our multi-task learning approach achieves higher efficiency compared to the baseline (MVNN) in mainly two settings: the 98-item scenario with a single kind of bidder and the 198-item setting with all the bidder population.
MT-MLCA-F gave a higher efficiency than MT-MLCA-R more often than the baseline.
Under such settings, several results of our methods were significantly higher than those of the baseline.
As shown in Figure \ref{fig:result_eff_hetero}, under the 196-item settings, our method maintains higher efficiency despite the increasing problem size, whereas the 98-item settings do not demonstrate such scalability.

\begin{figure*}[htb]
  \centering
  \includegraphics[width=0.8\linewidth]{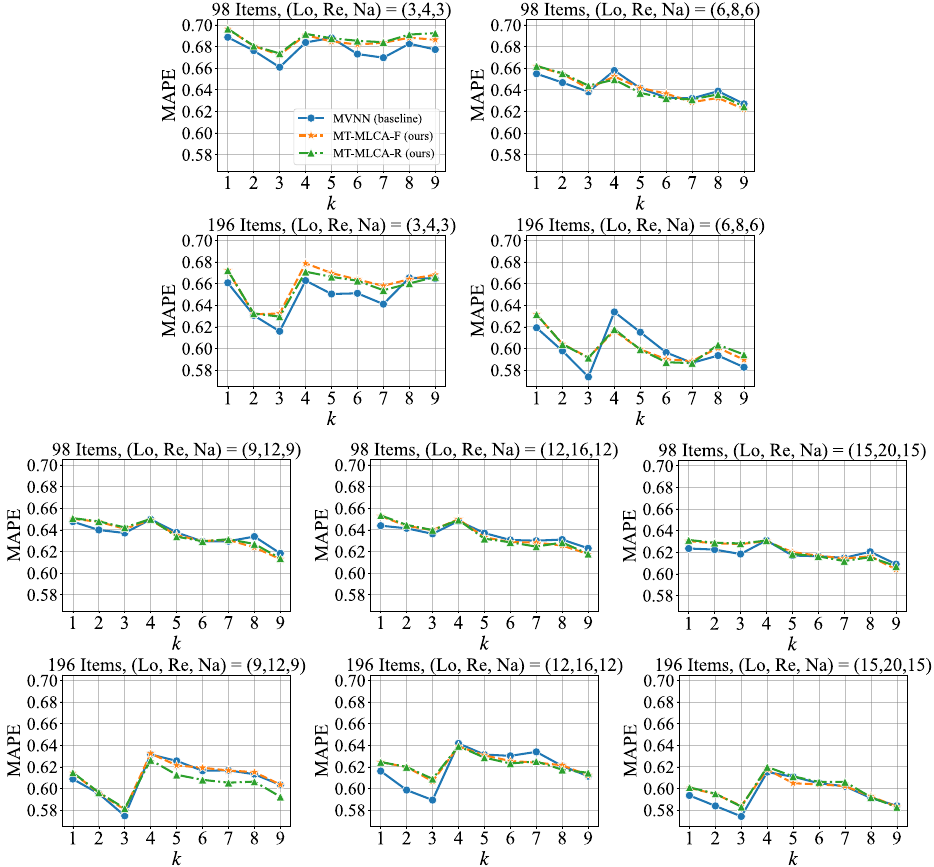}
  \caption{
    MAPE results for all settings.
    The horizontal axis represents the number of rounds $k$ in \textsc{MLCA}, and the vertical axis denotes the MAPE.
    The first row of five figures presents the results for 98 items with $(3l,4l,3l) (l=1,\dots, 5)$ bidders, while the second row of five figures displays the results for 196 items.
    All figures share the legend presented in the first one.
  }
  \label{fig:result_mape}
\end{figure*}

We show the variations in MAPE scores in Figure \ref{fig:result_mape}.
MAPE scores are calculated using the equation \eqref{eq:mape} for each round $k$
, which ranges from 1 to 9 ($= \lfloor(Q^\mathrm{max} - Q^\mathrm{round})/Q^\mathrm{round}\rfloor$).
In most large-scale scenarios ($(3l, 4l, 3l), l\ge 3$), our multi-tasking approach achieves lower MAPE by the final round ($k=9$).

\subsection{Discussion}

In this section, we discuss the experimental results and their implications for our multi-task learning approach in combinatorial auctions.

It is crucial to note that we did not perform hyper-parameter tuning and maintained uniform optimization methods across all bidders to ensure consistency across the experimental settings. 
This approach confirms that the enhancements in efficiency are due solely to our proposed multi-task learning techniques, not influenced by variability in model optimization. 
Furthermore, our approach achieved up to a 4\% increase in efficiency in scenarios with 196 items and 50 national bidders. 
Even this marginal improvement can translate into significant monetary benefits in real-world applications \cite{weissteiner22monotone}, emphasizing the practical value of these improvements.

From Table \ref{tab:result_eff}, we observe that our multi-tasking approach performs well in several settings, though it does not consistently outperform the baseline.
The efficiency results for 50 bidders of one kind suggest that multi-tasking can effectively capture intrinsic task relatedness through shared parameters, making it suitable for auction environments where bidders have similar bidding strategies.
However, under the auction environments with 98 items, MVNN outperforms our methods, while it underperforms in comparison to ours in the 196-item scenarios.
This discrepancy could be attributed to the task unrelatedness or relatedness resulting from the way we used to increase the number of items.
We doubled the number of default spectrum band blocks to create the environments with 196 items.
This increased substitutability among items, augmenting the relatedness of the valuation functions and, thus, the efficiency of multi-task learning.
The pool of relevant information for estimating valuations across different bidders also increases as the number of nearly equivalent bundles grows.
Therefore, our multi-tasking approach can more effectively utilize interconnected data for precise valuation estimations in 196-item settings compared to 98-item environments.
The MAPE results for large-scale auctions with 196 items and $(3l, 4l, 3l), (l\ge 3)$ bidders, as shown in Figure \ref{fig:result_mape}, corroborate this explanation because MAPE for 196 items are generally lower than those for 98 items.

MT-MLCA-F and MT-MLCA-R do not consistently exhibit superiority in both efficiency and MAPE.
This is due to the shallow architecture of the original MVNN. 
The original paper \cite{weissteiner22monotone} suggests the number of hidden layers at most 3; therefore, there is no obvious significance between MT-MLCA-F, which shares the feature extraction and differentiates the regression task, and MT-MLCA-R, which conversely operates the two process.

All methods, including ours, show lower efficiency around 0.4, due to the challenge of estimating bidders' true valuations with only 10 queries. 
Enhancing the efficiency in such scenarios is a goal for future work.


\section{Conclusion}
Our research introduces multi-task learning in iterative combinatorial auctions for preference elicitation in which many bidders exhibit similar valuation functions. 
We applied soft-parameter sharing to effectively estimate these interconnected valuations, differentiating tasks based on bidder ID features. 
This approach was more effective in our simulation-based experiments, particularly in scenarios involving a large number of bidders with similar bidding patterns and a substantial number of items, compared to existing methods.

In our experiments, we observed improvements in auction outcomes when employing our multi-task learning approach, while standardizing other hyper-parameters and optimization algorithms across all models. 
Specifically, our approach achieved higher efficiency than the existing method in settings with 196 items and over 30 bidders, as well as with either 98 or 196 items with over 50 bidders of the same kind. 
Additionally, our multi-tasking approach achieved lower prediction errors, as measured by MAPE scores, in large-scale scenarios. 
These results demonstrate that our multi-tasking approach can more effectively utilize interconnected data for precise valuation estimations for optimizing the efficiency, with potential increase in the total revenue for the auctioneer.
The observed improvements in these specific settings align with our research objectives and underscore the effectiveness of our approach within its intended context.

Enhancing the efficiency of our auction models in scenarios with sparse data will be a primary focus of our future research efforts.

\begin{credits}
\subsubsection{\ackname} This work is supported by JSPS KAKENHI Grant Number 21H04979.
\subsubsection{\discintname}
The authors have no competing interests to declare that are relevant to the content of this article. 
\end{credits}

%
%
%
\bibliographystyle{splncs04}
\bibliography{mybibfile}

\setcounter{section}{0} 
\renewcommand{\thesection}{Appendix \arabic{section}}
\section{Discussion on Identical Architecture}
\label{apdx:identical}
We assume that the machine learning models in \textsc{MLCA}, denoted as $\{\mathcal{A}_i\}_{i\in N}$, all share a common architecture. 
This assumption is elaborated on in the following discussion.

In auctions, bidders are characterized by their {\em types}, a form of private information that critically influences their bidding strategies and behaviors. 
Typically, types are identical to the true valuation functions, but all the information influencing the functions can be seen as constituents of types.
A prime example of this concept is evident in spectrum auctions, particularly as enterprise scales \cite{weiss17sats}.
In these auctions, the bidders are usually mobile network operators vying for licenses to use spectrum band blocks.
The diversity in bidder objectives is clear here: while some aim to acquire as many spectrum bands as possible, others might focus their interest on securing licenses for specific geographic regions.

Current methodologies, as shown in recent studies \cite{weissteiner22monotone,Weissteiner_Heiss_Siems_Seuken_2023}, often involve extensive hyper-parameter optimization to determine the optimal model structure for each bidder.
This optimization, applied to each model $\mathcal{A}_i$, considers the bidders' previously mentioned valuation tendencies. 
Essentially, this technique involves model tuning with knowledge of the bidder types.

In contrast to these existing methods, our approach operates under the assumption that bidder types are anonymous to the auctioneer. 
This decision is predicated on the practical limitation that personalizing models to individual bidders is not practicable in the absence of detailed information regarding bidder types.
Consequently, we utilize a uniform model structure for all bidders during the estimation phase. 
By applying a uniform model across all types, we ensure that the estimation process is not only fair and scalable but also remains robust against the variability of unknown bidder types, particularly in large-scale auctions where diverse bidder strategies are present.

\section{Discussion on Incentive Properties}
\label{apdx:incentive}
\textsc{MLCA} has been shown to ensure truthfulness, individual rationality, and no-deficit.
In this section, we describe how \textsc{MLCA} ensures these properties, and also how our approach inherits them.

First of all, \textsc{MLCA} inherently ensures no-deficit, which means that the monetary transfer between bidders and the auctioneer is non-negative. This is because the solution $A^*_{R_{-i}}$ makes the payment function \eqref{eq:vcgpayment} non-negative.
More formally, the payment for bidder $i$ is 
\begin{align*}
    &p(R)_i \\
    &= \sum_{j\in N\setminus\{i\}} \bar{v}_j(\bm{a}^*_{R_{-i},j}) - \sum_{j\in N\setminus\{i\}} \bar{v}_j(\bm{a}^*_{R,j})\\
    &= \left(\max_{A\in \mathcal{F}}\sum_{j\in N\setminus\{i\}: (\bm{a}_j, \bar{v}_j(\bm{a}_j))\in R_j} \bar{v}_j(\bm{a}_j)\right) - \sum_{j\in N\setminus\{i\}} \bar{v}_j(\bm{a}^*_{R,j}).
\end{align*}
Since $\bm{a}^*_{R,j}$ also satisfies $(\bm{a}^*_{R,j}, \bar{v}_j(\bm{a}_j))\in R_j$ for all $j\in N\setminus\{i\}$, the allocation $[\bm{a}^*_{R,1}~\cdots~\bm{a}^*_{R,n}]$ is also feasible for the first term. 
Therefore, the payment is non-negative by maximization.

Individual rationality, which means that bidders' utilities are non-negative, also applies as follows:
\begin{align}
    &u_i(\bm{a}^*_{R,i}, p(R)_i) \notag\\
    &= \underbrace{\left(\bar{v}_i(\bm{a}^*_{R,i}) + \sum_{j\in N\setminus\{i\}} \bar{v}_j(\bm{a}^*_{R,j})\right)}_\text{Reported SW of main economy} - \underbrace{\sum_{j\in N\setminus\{i\}} \bar{v}_j (\bm{a}^*_{R_{-i}, j})}_\text{Reported SW of marginal economy}\label{eq:decomposition_util}\\
    &= \max_{A\in \mathcal{F}}\left(\sum_{i\in N: (\bm{a}_i, \bar{v}_i(\bm{a}_i))\in R_i} \bar{v}_i(\bm{a}_i)\right) - \max_{A\in\mathcal{F}}\left(\sum_{j\in N\setminus\{i\}: (\bm{a}_j, \bar{v}_j(\bm{a}_j))\in R_j} \bar{v}_j(\bm{a}_j)\right)\notag\\
    &\ge 0.\notag
\end{align}
The last inequality holds because the solution for the second term is also feasible in the first term, and the bidder $i$'s report is assumed to be non-negative.

For the truthfulness, Brero et al.~\cite{brero2021machine} made an assumption regarding the above equations.
In equation \eqref{eq:decomposition_util}, the sum of the first two terms is called the reported social welfare (SW) of the main economy, and the last term is called the reported social welfare of (bidder $i$'s) marginal economy~\cite{brero2021machine}.
Brero et al.~\cite{brero2021machine} assumed that if all the bidders except $i$, the reported social welfare of $i$'s marginal economy is independent of the reports of bidder $i$.
The justification for this assumption is as follows: the query that optimizes the social welfare of this marginal economy is generated per round in Line \ref{algline:marginal_econ_8} of the Algorithm \ref{alg:mlca}, and thus almost all the bids that constitute $A^*_{R_{-i}}$ have been queried without considering the report of bidder~$i$.

In addition to this assumption, they assumed that if all bidders bid truthfully, then \textsc{MLCA} will find an efficient allocation. 
These two assumptions prove that truthful bidding is an ex post Nash equilibrium.
The idea is as follows.
Under the first assumption, the only way for bidder $i$ to increase his utility is to increase the reported social welfare of the main economy.
The latter assumption ensures that \textsc{MLCA} maximizes the social welfare if bidder $i$ reports truthfully.
These two arguments conclude that bidding truthfully is the best response for bidder $i$ if all bidders report truthfully.

Our modification, which uses multi-task learning in the estimation step, does not break these assumptions.
Multi-tasking increases the accuracy of the estimation and does not affect any part of the overall algorithm.
The increase in estimation accuracy does not affect the ability to find the efficient allocation, thus preserving the latter assumption.
In addition, this effect supports the justification for the first assumption: the increased accuracy must increase the likelihood that all bids that make up $A^*_{R_{-i}}$ are elicited with fewer queries.
Therefore, we do not break the first assumption.

\end{document}